\documentclass[pra,twocolumn,showkeys,showpacs,superscriptaddress,10pt]{revtex4-1}  
\usepackage{graphicx,epsfig,epstopdf}
\usepackage[dvipsnames]{xcolor}
\usepackage{amsmath}
\usepackage{subfigure}
\usepackage{mathrsfs}
\usepackage{bm}
\usepackage{times}

\begin{document}

\title{Interaction induced  non-reciprocal three-level quantum transport}

\author{Sai Li}

\author{Tao Chen}
\affiliation{Guangdong Provincial Key Laboratory of Quantum Engineering and Quantum Materials,
and School of Physics\\ and Telecommunication Engineering,
	South China Normal University, Guangzhou 510006, China}

\author{Jia Liu}\email{liuj.phys@foxmail.com}

\author{Zheng-Yuan Xue}\email{zyxue83@163.com}
\affiliation{Guangdong Provincial Key Laboratory of Quantum Engineering and Quantum Materials,
and School of Physics\\ and Telecommunication Engineering,
	South China Normal University, Guangzhou 510006, China}

\affiliation{Frontier Research Institute for Physics, South China Normal University, Guangzhou 510006, China}

\date{\today}

\begin{abstract}
Besides its fundamental importance,  non-reciprocity has also found many potential applications in quantum technology. Recently, many quantum systems have been proposed to realize non-reciprocity, but stable non-reciprocal process is still experimentally difficult in general, due to the needed cyclical interactions in artificial systems or operational difficulties in solid state materials. Here, we propose a new kind of interaction induced non-reciprocal operation, based on the conventional STIRAP setup, which removes the experimental difficulty of requiring cyclical interaction, and thus it is directly implementable in various quantum systems. Furthermore, we also illustrate our proposal on a chain of three coupled superconducting  transmons,  which can lead to a non-reciprocal circulator with high fidelity without a ring coupling configuration as in the previous schemes or implementations. Therefore, our protocol provides a promising way to explore fundamental non-reciprocal quantum physics as well as realize non-reciprocal quantum device.
\end{abstract}

\keywords{Non-reciprocity, quantum transport, superconducting quantum circuits}

\pacs{03.65.Ta, 03.65.Aa, 03.67.Lx}

\maketitle

\section{Introduction}
Reciprocity, which means that the measured scattering does not change when the source and the detector are interchanged \cite{Reci}, is a fundamental phenomenon in both classical and quantum regimes.  Meanwhile,  non-reciprocal devices, such as isolators and circulators, are also essential in both classical and quantum information processing. Especially, circulators can separate opposite signal flows, spanning from classical to quantum computation and communication systems \cite{QCir}. Thus, circulators are vital for  the design of full-duplex communication systems, which can transmit and receive signals through a same frequency channel, providing the opportunity to enhance channel capacity and reduce power consumption \cite{QCir2}. Therefore, many theoretical and experimental progresses have been made recently to build non-reciprocal devices in different quantum systems. Specifically, the conventional way of realizing non-reciprocity is achieved by adding magnetic field or using magnetic materials directly \cite{magnetic}. However, the external magnetic field would affect the transformation and magnetic materials is hardly to induce non-reciprocity.

Recently, realization of non-reciprocity has been proposed in many artificial quantum systems, such as in non-linear systems \cite{nonlinear1,nonlinear2,nonlinear3}, synthetic magnetism systems \cite{synthetic1,synthetic2,synthetic3,synthetic4,synthetic5, synthetic6,synthetic7}, non-Hermitian systems \cite{NH1,NH2,NH3,NH4,NH5,NH6,NH7,NH8}, time modulated systems \cite{TM1,TM2,TM3,TM4,TM5,TM6,TM7,TM8,TM9,TM10,TM11,NPtramsmon}, etc. Although these successful methods have been realized in nitrogen-vacancy centers systems \cite{synthetic7}, cold atom systems \cite{NH8}, superconducting circuits \cite{TM11, NPtramsmon},  and optical systems \cite{optical1,optical2,optical3}, tunable non-reciprocal process is still lacking for quantum manipulation. This is because previous schemes rely highly on special material properties, e.g., nonlinear property. In addition, for the experimental implementations of  non-reciprocity induced by synthetic magnetism, e.g., on superconducting circuits, they usually  need cyclical interaction among at least three levels of a superconducting qubit device \cite{NPtramsmon} or three coupled superconducting qubit devices in a two-dimensional configuration \cite{Nori-rew-Simu2-JC,TM11}. These are experimentally challenging for large scale lattices, as they require that  quantum systems have cyclical transition or at least two-dimensional configuration for  three-level  non-reciprocal process. Therefore, proposals using tunable non-reciprocal process and its potential applications to achieve non-reciprocity are still highly desired theoretically and experimentally.

Here, we propose a general scheme on a three-level quantum system based on the conventional STIRAP setup \cite{stirap1, stirap2} to realize non-reciprocal operations by time modulation. The distinct merit of our proposal is that the realization only needs two time-modulation coupling, which removes the experimental difficulty of requiring cyclical interaction, and thus it is directly implementable in various quantum systems, for example, superconducting quantum circuits systems \cite{QCir,Nori-rew-Simu2-JC,cqed2,cqed3}, nuclear magnetic resonance systems \cite{NMR}, a nitrogen-vacancy center in diamonds \cite{synthetic7}, trapped ions  \cite{Ion}, hybrid quantum systems \cite{HS}, and so on. Meanwhile, the three-level  non-reciprocal process can be implemented in a one-dimensional configuration instead of two-dimensional system in previous proposals, {which greatly releases the experiment difficulties for large lattices.

It is well known that  superconducting quantum circuits system is scalable and controllable, and thus attracts great attention in many researches.
Different from the cold atoms and optical lattice systems,
superconducting circuits possess good individual controllability and easy scalability. Following that, we illustrate our proposal on a chain of  three coupled  superconducting transmon   devices with appropriate parameters, and achieve a non-reciprocal circulator with high fidelity. As our proposal is based on a one-dimension superconducting lattice, e.g. Refs. \cite{sun2018,sun2019}, and with demonstrated techniques there, thus it can be directly verified.
Therefore, our proposal provides a new approach based on time modulation for engineering non-reciprocal devices, which can find many interesting  applications in quantum
information processing, including one-way propagation of quantum information, quantum measurement and readout, and quantum steering.

\begin{figure}[tbp]
	\begin{center}
		\includegraphics[width=7cm]{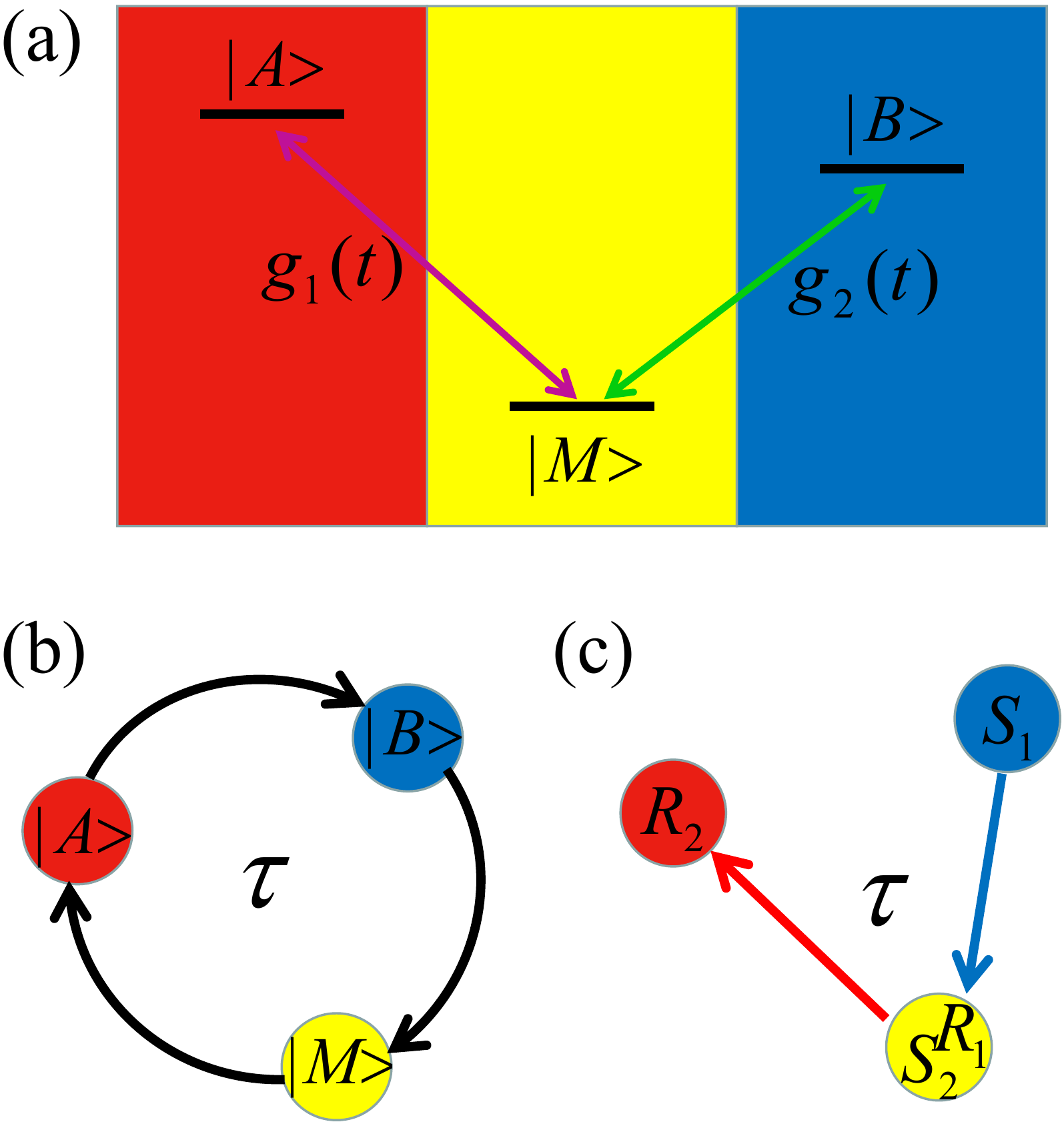}
\caption{Schematic diagram. (a)  Initial picture for generating non-reciprocal devices: two subspaces $|A\rangle$ and  $|B\rangle$ are resonantly coupled to subspace $|M\rangle$  simultaneously with time modulated coefficients $g_1(t)$ and $g_2(t)$. (b) Quantum circulator with one direction flow through a period $\tau$. (c) $|M\rangle$ simultaneously serve as receiver $R_1$ and sender $S_2$ for receiving  information from sender $S_1$ and send information to receiver $R_2$ through a period $\tau$.}\label{F1}
	\end{center}
\end{figure}

\section{General framework}
Now, we start from a general three-level quantum system labeled in the Hilbert space $\{|A\rangle, |M\rangle, |B\rangle\}$. As shown in Fig. \ref{F1}(a), considering  $|A\rangle$ and $|B\rangle$ simultaneously coupled  to $|M\rangle$ resonantly. Assuming $\hbar=1$ hereafter, the interaction Hamiltonian in the interaction picture can be written as
\begin{eqnarray}\label{HI}
  \mathcal{H}(t) &=& g_1(t) |A\rangle\langle M| + g_2(t)  |B\rangle\langle M| + \textrm{H.c.},
\end{eqnarray}
where  $g_{1,2}(t)$ are the time-modulation coupling strength.  Based on the Hamiltonian $\mathcal{H}(t) $, even if there is no direct coupling between bare states $|A\rangle $ and  $|B\rangle$, these two bare states are both coupled to state $ |M\rangle$, then, the transition between bare states $|A\rangle $ and  $|B\rangle$ can be realized via the middle state $|M\rangle$, i.e. STIRAP  \cite{stirap1, stirap2}. Especially, when the pulse shapes of $g_1(t)$ and $ g_2(t)$ are different from each other, the symmetry of the system can be broken naturally. Thus, with appropriate designed time-modulation coupling strengths $g_{1,2}(t)$, a quantum circulator with one-direction flow can be achieved through a period evolution with time $\tau$, i.e., $|A\rangle$ $\rightarrow$ $|B\rangle$$\rightarrow$  $|M\rangle$ $\rightarrow$  $|A\rangle$ illustrated in in Fig. \ref{F1}(b). That means, on the one hand, transition $|A\rangle$ $\rightarrow$ $|B\rangle$ is allowed, not vice versa. Meanwhile, the process $|B\rangle$ $\rightarrow$  $|M\rangle$ and $|M\rangle$ $\rightarrow$  $|A\rangle$ can be simultaneously realized, which means that $|M\rangle$ can simultaneously receive the quantum information from sender $|B\rangle$ and send information to receiver $|A\rangle$.  Those quantum processes are very important for quantum information transformation and processing.

\section{Construction}

Here, we proceed to introduce details for construction of non-reciprocal scattering matrix. Generally, the evolution process of time-dependent Hamiltonian with period $\tau$ can be expressed as $U(\tau) = \mathcal{T}\exp[-i\int^\tau_0 \mathcal{H}(t)dt]$, where $\mathcal{T}$ is time-order operator. When the quantum system satisfies the von-Neumann equation ${\partial}I(t)/{\partial t}+i[\mathcal{H}(t),I(t)]=0$ with dynamical invariant $I(t)$, the concrete solution of the evolution process with time-dependent Hamiltonian $\mathcal{H}(t)$ can be carried out by Lewis-Riesenfeld (LR) invariant method \cite{LR}. Here, we concentrate on the solution of nonreciprocal transition evolution process with time modulation. With the LR method, the evolution process with period $\tau$ can be exactly derived as
\begin{eqnarray}\label{UT}
  U(\tau) &=& \sum_{n=0,\pm} e^{-i\theta_n(\tau)}|\mu_n(\tau)\rangle\langle\mu_n(0)|,
\end{eqnarray}
 where  in the Hilbert space $\{|A\rangle, |M\rangle, |B\rangle\}$,
 \begin {equation}\label{eigen1}
\left|\mu_{0}(t)\right\rangle=\left[ \begin{array}{c}{\cos \gamma(t) \cos \beta(t)} \\ {-i \sin \gamma(t)} \\ {-\cos \gamma(t) \sin \beta(t)}\end{array}\right]
 \end {equation}
 and
 \begin {equation}\label{eigen2}
\left|\mu_{ \pm}(t)\right\rangle=\frac{1}{\sqrt{2}} \left[ \begin{array}{c}{\sin \gamma(t) \cos \beta(t) \pm i \sin \beta(t)} \\ {i \cos \gamma(t)} \\ {-\sin \gamma(t) \sin \beta(t) \pm i \cos \beta(t) }\end{array}\right]
 \end {equation}
 are the eigenstates of the invariant \cite{LR1}
 \begin{equation}\label{a4}
I(t)=\frac{\mu}{2}
 \left(
   \begin{array}{ccc}
    0 & \cos{\gamma}\sin{\beta} & -i\sin{\gamma} \\
     \cos{\gamma}\sin{\beta} & 0 &  \cos{\gamma}\cos{\beta}\\
     i\sin{\gamma} & \cos{\gamma}\cos{\beta} &  0 \\
   \end{array}
 \right),
\end{equation}
where $\mu$ is an arbitrary constant with unit of frequency to keep $I(t)$ with dimensions of energy,  $\gamma(t)$ and $\beta(t)$ are auxiliary parameters , which satisfy the von-Neumann equation ${\partial}I(t)/{\partial t}+i[\mathcal{H}(t),I(t)]=0$, and $\theta_n(\tau)$ is the LR phase with $\theta_0(\tau)=0$ and $\theta_-(\tau)=-\theta_+(\tau)$, which can be addressed by auxiliary parameters $\gamma(t)$ and $\beta(t)$. To induce non-reciprocal transition evolution process, we set the boundary conditions as
\begin{eqnarray}\label{initial}
\gamma(0)= \gamma(\tau) = 0,  \quad  \beta(0) = 0, \quad \beta(\tau) = \pi/2.
\end{eqnarray}
After that, the final evolution operator in the Hilbert space $\{|A\rangle,|M\rangle,|B\rangle\}$ can be determined as
 \begin {equation}\label{cir}
U[\theta_+(\tau)]=\left[ \begin{array}{ccc}{0} & -i\sin{\theta_+(\tau)} & \cos{\theta_+(\tau)} \\ 0 & {\cos{\theta_+(\tau)}} & -i\sin{\theta_+(\tau)} \\ -1 & 0 & {0}\end{array}\right].
 \end{equation}

\begin{figure}[tbp]
	\begin{center}
		\includegraphics[width=6.5cm]{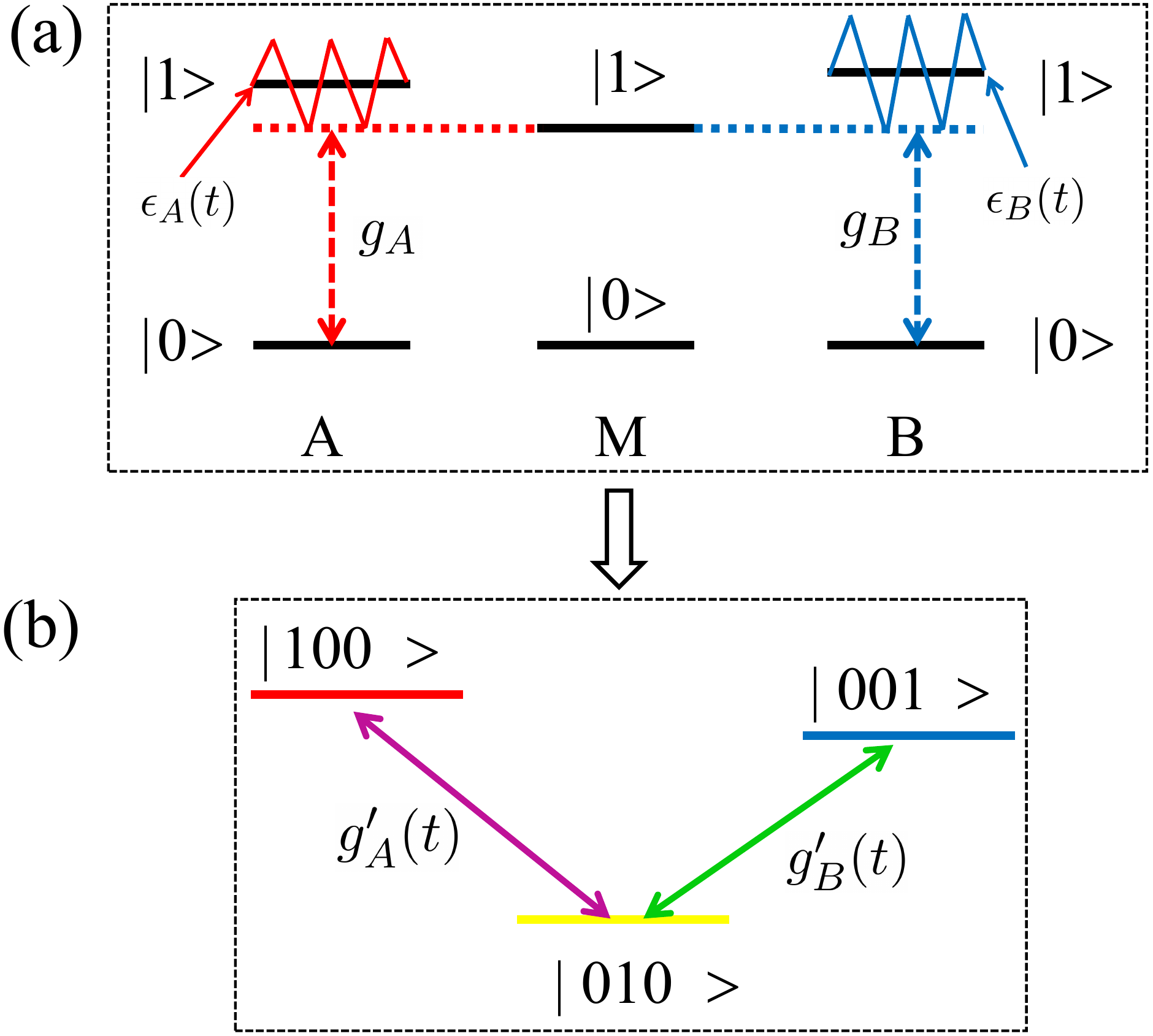}
\caption{Illustration of our scheme with three transmon devices. (a) Two qubit-frequency driven transmons $A$ and $B$ with the respective longitudinal field  $\epsilon_{A,B}(t)$  resonantly coupled to the transmon $M$. (b) Effective resonant coupling architecture in the single-excitation subspace $\{|100\rangle,|010\rangle,|001\rangle\}.$ }\label{F2}
	\end{center}
\end{figure}

\begin{figure*}[tb]
	\begin{center}
		\includegraphics[width=0.7\linewidth]{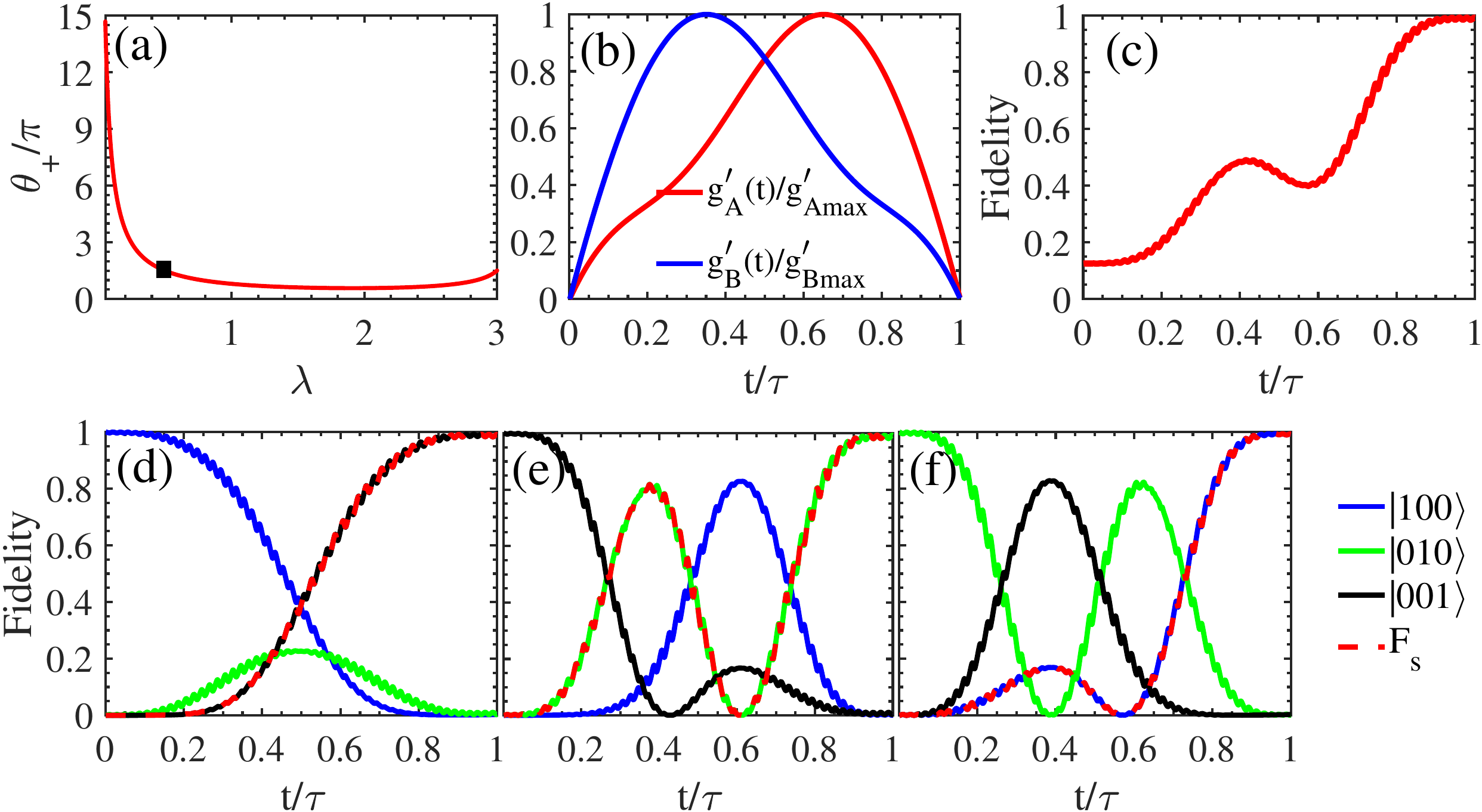}
\caption{Numerical performance. (a) The LR phase $\theta_+(\tau)$ with respect to auxiliary parameter $\lambda$, where black square represents $\theta_+(\tau)=3\pi/2$. (b) The pulse shapes of asymmetrical time-modulation coupling strength $g^\prime_{A,B}(t)$ with $\theta_+(\tau)=3\pi/2$ in a period $\tau$. (c) The fidelity of the quantum circulator $U[3\pi/2]$ for simultaneously sending and receiving the quantum information in a period $\tau$. (d), (e) and (f) The state populations and fidelity of the quantum circulator $U[3\pi/2]$ in a period $\tau$ with the initial state being $|100\rangle$, $|001\rangle$ and $|010\rangle$, respectively. 
}\label{F3}
	\end{center}
\end{figure*}

To understand the result clearly, for the case $ \theta_+(\tau) = \pi $, the final evolution operator $U[\pi]=-|M\rangle\langle M| -|A\rangle\langle B|-|B\rangle\langle A| $ represents a normal two-direction transitions. Especially,  for another case $ \theta_+(\tau) = 3\pi/2 $, the evolution process $U[3\pi/2]=i|A\rangle\langle M| +i|M\rangle\langle B|-|B\rangle\langle A| $  shows non-reciprocal transitions, that means transition $|A\rangle$ $\rightarrow $ $|B\rangle$ is allowed and transition $|B\rangle$ $\rightarrow $ $|A\rangle$ is forbidden for the same process. Obviously, the evolution operator $U[3\pi/2]$ also means a cyclic transportation. {To sum up, the evolution operator $U[3\pi/2]$ induces a cyclic chiral transportation,  which exactly realizes a quantum circulator, from pure time-modulation of the interaction.}

\section{Illustrative scheme with transmons}

Here, we propose a scheme on superconducting quantum circuits. For a transmon \cite{Tra}, there are three lowest levels, which can be resonantly driven by two microwave fields to induce the Hamiltonian $\mathcal{H}(t)$ in Eq. (\ref{HI}) our proposal. {In this case, only  nonreciprocal state transfer within a transmon can be obtained, and the implementation is straightforward, i.e., letting $|0\rangle$, $|1\rangle$, and $|2\rangle$ take the role of $|A\rangle$, $|M\rangle$, and $|B\rangle$.}

Furthermore, we consider a {more interesting case, that is,} three coupled transmons implementation, with the lowest two levels $|0\rangle$  and $|1\rangle$  in superconducting quantum circuits. As shown in Fig. \ref{F2}(a), we labeled three transmons with $A$, $M$ and $B$ with frequencies $\omega_{A,M,B}$ and anharmonicities $\alpha_{A,M,B}$. Here, we introduce qubit frequency drives $f(\epsilon(t))$ \cite{TM}, which can be determined experimentally by the
longitudinal field $\epsilon(t) = f^{-1}(\dot{F}(t))$, where $F(t)=\eta(t)\sin(\nu t)$ is intentionally chosen with $\nu$ being the frequency of the longitudinal field $\epsilon(t)$, and two qubit-frequency drives $f(\epsilon_{j}(t))$ $(j=A,B)$ are added in transmons $A$ and $B$ respectively to induce time-modulation resonant interaction with transmon $M$. Then, the coupled system can be described by $\mathcal{H}_T(t)= \mathcal{H}_f(t)+\mathcal{H}_{int}(t)$, where
$\mathcal{H}_f(t)$ and $\mathcal{H}_{int}(t)$ are free and interaction Hamiltonian respectively. For the free part,
\begin{equation}\label{B1}
\mathcal{H}_f(t)=\sum_{j=A,B}\{[\omega_j+f(\epsilon_j(t))]|1\rangle_j\langle1|\}
+\omega_M|1\rangle_M\langle1|,
\end{equation}
where $\epsilon(t) = f^{-1}(\dot{{F}}(t))$ with ${F}(t) = \eta(t)\sin(\nu t )$. For the interaction term,
\begin{equation}\label{B2}
\mathcal{H}_{int}(t)=[\sum_{j=A,B}g_j(|0\rangle_j\langle1|+\textrm{H.c.})]
\cdot(|0\rangle_M\langle1|+\textrm{H.c.}).
\end{equation}
Transforming to the rotating frame defined by $V=V_1+V_2$, where
$V_1 =\exp[i(\sum_{j=A,B}\omega_j|1\rangle_j\langle1|
+\omega_M|1\rangle_M\langle1|)]$
and
$V_2=\exp[i\sum_{j=A,B}{F}_j(t)|1\rangle_j\langle1|]$,
and the transformed Hamiltonian is
\begin{eqnarray}\label{B5}
\mathcal{H}_t(t)&=&V^\dagger \mathcal{H}_T(t)V+i\frac{dV\dagger}{dt}V\notag\\
&=&\sum_{j=A,B}g_j[|0\rangle_j\langle1|e^{-i\omega_jt}e^{i{F}_j(t)}+\textrm{H.c.}]\\
&\otimes&[|0\rangle_M\langle1|e^{-i\omega_Mt}+\textrm{H.c.}],\notag
\end{eqnarray}
in the single-excitation subspace $\{|100\rangle,|010\rangle,|001\rangle\}$, where $|amb\rangle\equiv|a\rangle_A\otimes|m\rangle_M\otimes|b\rangle_B$ labels the product states of three transmons, after neglecting the high order oscillating  terms, the Hamiltonian can be written as
\begin{eqnarray}\label{Hint}
\mathcal{H}_I(t)=&&g_A|100\rangle\langle010|e^{i\Delta_A t -iF_A(t)}+ \notag\\ &&g_B|001\rangle\langle010|e^{i\Delta_B t -iF_B(t)}+\textrm{H.c.},
\end{eqnarray}
where $g_{j}$ is coupling strength for transmons $A, B$ to $M$ and $\Delta_{j}=\omega_{j}-\omega_M$ are the frequency difference. Then, using the Jacobi-Anger identity expansion
$\exp(iF_j(t))=\sum^\infty_{m=-\infty}i^mJ_m[\eta_j(t)]\exp[im(\nu_jt-\frac{\pi}{2})]$, 
and considering the  resonant interaction case $\Delta_j = \nu_j$, the effective  Hamiltonian with time modulation can be obtained as
\begin{equation}\label{Hef}
\mathcal{H}_{\textrm{eff}}(t)=\frac{g^\prime_A(t)}{2}|100\rangle\langle010| +\frac{g^\prime_B(t)}{2}|001\rangle\langle010|+\textrm{H.c.},
\end{equation}
where $g^\prime_{j}(t) = 2g_{j}J_1(\eta_{j}(t)) $ are effective time-modulation coupling strength for transmons $A, B$ to $M$ with $J_1$ being the Bessel function. We can use the effective Hamiltonian $\mathcal{H}_{\textrm{eff}}(t)$ to realize quantum circulator in our protocol.

With the LR invariant method \cite{LR1}, according to the von-Neumann equation ${\partial}I(t)/{\partial t}+i[\mathcal{H}_{eff}(t),I(t)]=0$, the form of $g^\prime_{j}(t)$ can be given as
\begin {eqnarray}\label{relationship}
g^\prime_A(t)=2[\dot{\beta}(t) \cot \gamma(t) \sin \beta(t)+\dot{\gamma} (t)\cos \beta(t)],\notag\\
g^\prime_B(t)=2[\dot{\beta}(t) \cot \gamma(t) \cos \beta(t)-\dot{\gamma}(t) \sin \beta(t)].
 \end {eqnarray}
Considering the boundary conditions Eq. (\ref{initial}), the commutation relations $[H(0),I(0)]=[H(\tau),I(\tau)]=0$, and the experimental apparatus restriction, the values $g^\prime_{j}(t)$ can be set as zeros at time $t = 0$ and $\tau$, thus, a set of auxiliary parameters $\gamma(t)$ and $\beta(t)$ can be selected in a proper form \cite{LR2} as
\begin {eqnarray}\label{GB}
\gamma(t) &=& \frac{\lambda}{(\tau/2)^4} t^2(t-\tau)^2 ,\notag\\
\beta(t) &=& \dfrac{-10\pi t^7}{\tau^7}+\dfrac{35\pi t^6}{\tau^6}-\dfrac{42\pi t^5}{\tau^5}+\dfrac{35\pi t^4}{2\tau^4},
\end {eqnarray}
where $\lambda$ is a tunable time-independent auxiliary parameter, which directly determines the LR phase $\theta_+(\tau)$ concerned in our proposal shown in Fig. \ref{F3}(a). Furthermore, the effective coupling strength $g^\prime_{j}(t)$ can be carried out according to Eq. (\ref{relationship}). Then, we realize the final evolution operator $U[\theta_+(\tau)]$.

Following, we choose appropriate experimental parameters \cite{Martinis14} and show how to realize our protocol to achieve non-reciprocal operations on superconducting quantum circuits. {Due to the anharmonicity of transmon qubits are relatively small, thus the second excited state will contribute harmfully for the quantum process. To numerically quantify this effect, we} set the  anharmonicity of three transmons as $\alpha_A = 2\pi\times220$  MHz, $\alpha_M = 2\pi\times 210  $  MHz and $\alpha_B = 2\pi\times 230 $  MHz. Meanwhile,
we set the frequency of the longitudinal field $\nu_j$ equal to the corresponding frequency difference $\Delta_j$ as $\nu_A = \Delta_A = 2\pi\times 345$  MHz and  $\nu_B =\Delta_B = 2\pi\times 345 $  MHz respectively to induce time-modulation resonant interaction in the single-excitation subspace. Furthermore, we set the decoherence rates of the transmons as $\Gamma_A = 2\pi\times 3$ kHz, $\Gamma_M = 2\pi\times 4 $ kHz and $\Gamma_B = 2\pi\times 5$ kHz, coupling strength for transmons $A, B$ to $M$ as $g_A=g_B=2\pi\times 10$ MHz and the quantum evolution period $\tau =  145$ ns. Then, to realize the quantum circulator $U[3\pi/2]$, we modify auxiliary parameter $\lambda=0.4974$ to make $\theta_+(\tau) = 3\pi/2$ and naturally determine the time-modulation coupling strength $g^\prime_{j}(t)$, whose pulse shapes are plotted in Fig. \ref{F3}(b), which is smooth and easily experimentally realized.

We numerically simulate the performance of the quantum circulator $U[3\pi/2]$  by using Lindblad master equation as
\begin{equation}
\dot{\rho}=i \left[\rho, \mathcal{H}_{I}(t)\right]+ \sum_{k=A,M,B}\Gamma_{k} \mathcal{L}\left(\hat{O}_{k}\right),
\end{equation}
where $\rho$ is the density matrix of the considered system and $\mathcal{L}(\hat{O}_{k})= \hat{O}_{k}\rho_{1} \hat{O}_{k}^{\dagger} -\hat{O}_{k}^{\dagger} \hat{O}_{k} \rho_{1}/2-\rho_{1} \hat{O}_{k}^{\dagger} \hat{O}_{k}/2$ is the Lindbladian of the operator $\hat{O}_{k}$ with $\hat{O}_{k} = |0\rangle_k\langle1|+ (|0\rangle_k\langle0|-|1\rangle_k\langle1|)$. We first evaluate the quantum circulator $U[3\pi/2]$ with initial state $|\psi_{I1}\rangle$ sequentially prepared on the states $|100\rangle$, $|001\rangle$ and $|010\rangle$. Define the state fidelity $F_s = \langle\psi_{T1}|\rho|\psi_{T1}\rangle $ , where $\rho$ is the density matrix and sequential target states $|\psi_{T1}\rangle$ are $-|001\rangle$, $i|010\rangle$ and $i|100\rangle$. As shown in Fig. \ref{F3} (d), (e) and (f), the sequential state fidelity $F_s$ are $99.08\%$, $99.25\%$ and $99.28\%$, thus, the results demonstrate the construction of  our quantum circulator. {In our simulation, we do not include  control errors of the external driving field, due to the following. First, they can be well-controlled experimentally. Second, our protocol is robust against the  control errors, as show in Fig. \ref{F3} (c) (d), (e) and (f), i.e., the target state populations are nearly flat around the final times.}

Then, we evaluate the ability to simultaneously send and receive the quantum states for the quantum circulator $U[3\pi/2]$. We set transmon M as both sender and receiver, which can send quantum information to transmon A and simultaneously receive different quantum information from transmon B. To demonstrate this, we set the initial states $|\psi_{I2}\rangle = \cos \vartheta |010\rangle + \sin \vartheta |001\rangle$, and the corresponding target states are $|\psi_{T2}\rangle= i\cos \vartheta |100\rangle + i\sin \vartheta |010\rangle$. We define the fidelity $F_m = \frac{1}{2\pi}{\int^{2\pi}_0}\langle\psi_{T2}|\rho|\psi_{T2}\rangle d\vartheta$ with the integration numerically performed for 1001 input states with $\vartheta$ being uniformly distributed over $[0,2\pi]$. As shown in Fig. \ref{F3}(c), the initial fidelity is less than 0.2 due to the big deference between initial states and target states; the non-monotonic behavior in the time evolution of the fidelity means that the evolution process is a non-reciprocal process; and we get the fidelity $F_m=99.23\%$, which shows the power of our scheme can realize an effective quantum transfer station  which can receive and send different quantum information in one direction; the infidelity is caused by decoherence about $0.52\%$ and leakage error about $0.25\%$.

\section{Conclusion}
In summary, we propose a general scheme based on time modulation to realize non-reciprocal operations. Our proposal can be easily realized in many quantum systems. We {illustrate} our proposal on superconducting quantum circuits  with two driving transmons simultaneously coupled to the middle transmon. Considering the scalability and controllability of the superconducting quantum circuits, our scheme provides promising candidates for non-reciprocal  quantum information processing and devices in the near future.

\section*{Acknowledgements}
This work was supported by 
the National Natural Science Foundation of China (Grant No.~11874156 and No.~11904111)
and the Project funded by China Postdoctoral Science Foundation (Grant No.~2019M652684).

\end{document}